\documentclass[a4paper,11pt]{article}
\usepackage{jinstpub} 
\usepackage{lineno}

\usepackage{graphicx}
\graphicspath{{figures/}}
\usepackage{dcolumn}
\usepackage{bm}
\usepackage{xspace}
\usepackage{placeins}
\usepackage[pdftex,table]{xcolor}
\definecolor{darkpurple}{rgb}{0.5, 0.2, 0.8}
\definecolor{darkblue}{rgb}{0.0, 0.0, 0.8}
\definecolor{darkgreen}{rgb}{0.0, 0.4, 0.0}
\definecolor{darkred}{rgb}{0.5, 0.0, 0.0}
\usepackage{hyperref}
\hypersetup{
    linktocpage,
     colorlinks,
     citecolor=darkgreen,
     linkcolor= darkgreen,
     urlcolor=darkgreen
}

\newcommand{\et}{\ensuremath{E_\text{T}}\xspace}
\newcommand{\cnf}{\ensuremath{\mathcal{CNF}}\xspace}

\newcommand{\dx}{\ensuremath{\Delta_{X}}\xspace}
\newcommand{\xt}{\ensuremath{{X}_{\text{true}}}\xspace}
\newcommand{\xr}{\ensuremath{{X}_{\text{reco}}}\xspace}

\newcommand{\m}{$\mu$\xspace}
\newcommand{\Hgg}{$H \rightarrow \gamma\gamma$}
\newcommand{\Hmm}{$H \rightarrow \mu\mu$}

\begin{document}

\title{Generative Machine Learning for Detector Response Modeling with a Conditional Normalizing Flow}
\author[1]{Allison Xu}
\emailAdd{allisonxu@berkeley.edu}

\author[2]{Shuo Han}
\emailAdd{shuohan@lbl.gov}

\author[3]{Xiangyang Ju}
\emailAdd{xju@lbl.gov}

\author[2,4]{Haichen Wang}
\emailAdd{haichenwang@berkeley.edu}
\affiliation[1]{Department of Statistics, University of California, Berkeley, Berkeley, CA 94720, USA}
\affiliation[2]{Physics Division, Lawrence Berkeley National Laboratory, Berkeley, CA 94720, USA}
\affiliation[3]{Scientific Data Division, Lawrence Berkeley National Laboratory, Berkeley, CA 94720, USA}
\affiliation[4]{Department of Physics, University of California, Berkeley, Berkeley, CA 94720, USA}

\date{\today}

\abstract{
In this paper, we explore the potential of generative machine learning models as an alternative to the computationally expensive Monte Carlo (MC) simulations commonly used by the Large Hadron Collider (LHC) experiments. Our objective is to develop a generative model capable of efficiently simulating detector responses for specific particle observables, focusing on the correlations between detector responses of different particles in the same event and accommodating asymmetric detector responses. We present a conditional normalizing flow model (\cnf) based on a chain of Masked Autoregressive Flows, which effectively incorporates conditional variables and models high-dimensional density distributions. We assess the performance of the \cnf model using a simulated sample of Higgs boson decaying to diphoton events at the LHC. We create reconstruction-level observables using a smearing technique. We show that conditional normalizing flows can accurately model complex detector responses and their correlation. This method can potentially reduce the computational burden associated with generating large numbers of simulated events while ensuring that the generated events meet the requirements for data analyses. We make our code available at \href{https://github.com/allixu/normalizing_flow_for_detector_response}{\texttt{https://github.com/allixu/normalizing\_flow\_for\_detector\_response}}}


\maketitle


\section{\label{sec:intro}Introduction}
The Monte Carlo (MC) simulation frameworks utilized by the Large Hadron Collider (LHC) experiments~\cite{ATLAS:2010arf,CMS:2006myw,LHCb:2005ab} play a crucial role in the success of its physics program, which probes physics beyond the Standard Model through precision measurements and direct searches. These MC simulation frameworks have been extensively tuned to model particle collisions and detector effects. In general, a simulation framework used by an LHC experiment is a chain of multiple components, including event generation, detector simulation, and event reconstruction. Each of these components may be further factorized into more focused tasks, which are primarily first-principle based, simulating the physics process or detector response according to our best theoretical and phenomenological knowledge of the collision process and detector material, respectively.  However, the simulation of MC samples, especially the modeling of detector response, is computationally expensive. As the LHC continues to operate successfully, particularly with its upcoming high luminosity program, existing simulation schemes face difficulties in meeting the computational demands that come with the significant increase in integrated luminosity.

The application of generative machine learning as a surrogate for certain aspects or the entirety of the Monte Carlo (MC) simulation utilized at the LHC is a promising solution actively being investigated by the high energy physics community. A significant area of development is the use of generative machine learning to model particle shower development in detectors~\cite{Paganini:2017dwg,Erdmann:2018jxd,atlfast3,Buhmann2020,AbhishekAbhishek:2022wby,Krause:2022jna,Krause:2021ilc,Mikuni:2022xry}. Recently, the ATLAS experiment at the LHC has incorporated a Generative Adversarial Networks (GAN) based fast calorimeter shower simulation into its fast detector simulation framework~\cite{SIMU-2018-04}. Another active area of investigation is the use of generative machine learning to model the collision, parton showering, hadronization, and jet formation processes~\cite{Butter:2019cae,Touranakou:2022qrp,Buhmann:2023pmh,DiSipio:2019imz,Otten:2019hhl,Stienen:2020gns,Choi:2021sku}. In terms of choice of machine learning architecture, Ref.s~\cite{Butter:2019cae,Paganini:2017dwg,Erdmann:2018jxd,SIMU-2018-04,DiSipio:2019imz,Otten:2019hhl,Choi:2021sku,Hashemi:2023ruu} utilized  
Generative Adversarial Networks (GAN), Ref.s~\cite{Buhmann2020,AbhishekAbhishek:2022wby} adopted autoencoders,  Ref.s~\cite{Diefenbacher:2023vsw,Krause:2022jna,Krause:2021ilc,Stienen:2020gns,10.21468/SciPostPhys.9.5.074} exploited normalizing flows, and Ref.~\cite{Mikuni:2022xry} explored diffusion models. More detailed reviews of the state of the art of generative machine learning for particle physics can be found in Ref.s~\cite{Butter:2020tvl,Butter:2022rso}.

In this paper, we target a different use case of generative machine learning. Many data analyses, targeting specific signatures, often do not need the detailed information of the collision final state produced from the full simulation framework. For example, in ATLAS \Hgg\ and \Hmm\ measurements, high-statistics background samples are generated for background modeling, and the equivalent integrated luminosity of these samples can be as large as 30 ab$^{-1}$~\cite{HIGG-2019-14,HIGG-2020-16}. In addition, as the Higgs boson measurements enter a precision phase, many analyses would require the simulation of a large number of signal samples with alternative physics parameters such as those defined in the Standard Model effective field theory, which is used in interpreting the observed results. Deploying the full simulation chain that uses GEANT4 package~\cite{GEANT4:2002zbu} to simulate detailed interactions between particles and detector materials is often unnecessarily inefficient and in some cases unrealistic, for such tasks. 

A generative machine learning model that inputs generator-level particle variables and generates the detector responses for specified particle observables is all we need for this kind of analysis use case. We identified the following design objectives: the model should learn the detector response to a given observable as a function of conditional variables; the model should learn the correlation between detector responses of different particles in the same event; and the model should learn asymmetric detector response, which is commonplace in particle detection. Some recent works~\cite{DiBello:2022rss} explored similar objectives using generative models incorporating novel attention mechanisms. In our work, we designed a conditional normalizing flow model (\cnf) to achieve these objectives. 
The \cnf model 
is based on a chain of Masked Autoregressive Flows~\cite{PapamakariosMAF}, which combines the advantages from the normalizing flow~\cite{pmlr-v37-rezende15} and the autoregressive density estimation~\cite{JMLR:v17:16-272}. The \cnf model can naturally include conditional variables and model high dimensional density distributions. 
This work parallels that of Ref.~\cite{10.21468/SciPostPhys.9.5.074} in utilizing an invertible normalizing flow architecture; however, our objectives differ. While this study focuses on generating detector responses, Ref.~\cite{10.21468/SciPostPhys.9.5.074} aims to unfold detector-level observables back to the parton-level scattering amplitude. Compared to fast simulation approaches based on smearing techniques, such as those used in DELPHES~\cite{Ovyn:2009tx}, the generative machine learning approach offers a straightforward method for modeling complex detector responses. This is particularly useful when the response shapes cannot be easily described by analytical functions or when there are correlations between detector responses.

We characterized the performance of the \cnf model using a simulated sample of Higgs boson decaying to diphoton (\Hgg) events at the LHC. For this sample, we engineered various physics-motivated detector response scenarios and created reconstruction-level observables using a smearing technique similar to that adopted by the fast detector simulation package DELPHES~\cite{Ovyn:2009tx}. 
This paper is organized as follows: Section~\ref{sec:model} reviews conditional normalizing flows; Section~\ref{sec:sample} describes the event generation and the smearing technique used to introduce experimental effects; Section~\ref{sec:architecture} presents the architecture and training configuration of our conditional normalizing flows model in more detail, and Section~\ref{sec:results} shows the performance of the \cnf\ tool in various scenarios; Section~\ref{sec:discussion} summarizes the findings and discusses potential applications and extensions of this tool as well as interesting future directions. Section~\ref{sec:conclusion}
concludes the paper.

\section{\label{sec:model}Conditional Normalizing Flows}

A normalizing flow is a technique that transforms a simple base density distribution $\pi(\vec{z})$ to a more complex target density distribution $p(\vec{x})$ using a bijective, differentiable function known as a bijection $\vec{x} = f(\vec{z})$. In an application of the normalizing flow, the $\vec{x}$ would be features to learn and generate, while $\vec{z}$ are random variables generated by a base density distribution. A normalizing flow often uses a chain of bijections to construct the final bijection, which allows the modeling of complex target distributions. To make the normalizing flow learnable and computationally efficient, a bijection is often chosen to be a simple function and the coefficients of the function are parameterized by neural networks, often by the MultiLayer Perceptrons (MLPs). 
Applying the change of variables method, as described in Equation \ref{eq:change_of_variable} , a normalizing flow can estimate the target density distribution with the input vector $\vec{x}$. The learnable weights in the neural network $\vec{w}$ are then optimized by minimizing the negative log-likelihood function  $\mathcal{L}(\vec{w} | \vec{x}) = -\mathop{{}\mathbb{E}}_{x}[\log p_w(\vec{x})]$. A normalizing flow can be extended to a conditional normalizing flow by concatenating the conditional vector $\vec{c}$ with the input vector $\vec{x}$ and using the combined vector to estimate the target density distribution. 

\begin{gather}
    p(\vec{x}) = \pi(\vec{z}) |\det{J_f(\vec{z})}|^{-1} = \pi(f^{-1}(\vec{x})) |\det{J_{f^{-1}}(\vec{x})}|\\
    \nonumber \text{where $J_f(\vec{z})$ is the Jacobian matrix of the function $f$ with respect to $\vec{z}$}
    \label{eq:change_of_variable}
\end{gather} 

 Our \cnf\ implementation was based on a type of normalizing flows, known as the Masked Autoregressive Flows (MAF). In MAF, the bijection transforms the base density distribution by sequentially transforming each dimension based on the previously transformed dimensions. This autoregressive feature transforms depending on the ordering of the input vector and is slow for sampling. To minimize the ordering effect, we added a permutation bijection to each MAF.

In this work, we achieve the generation of detector responses that vary as functions of particle kinematics and event conditions with a conditional normalizing flow model. The target density distribution $p(\vec{x})$ is a multidimensional distribution that describes the detector responses of particle kinematic observables $X$ and their correlation. The conditional vector comprises particle kinematics and event variables on which target detector responses depend.

\section{\label{sec:sample}Data Samples}

\subsection{Event Generation}

This study simulated the Higgs boson production in $pp$ collisions at $\sqrt{s} = 13$ TeV. The Higgs boson subsequently decays into a pair of photons. The events were generated by the Madgraph@NLO (v2.3.7)~\cite{Alwall:2014hca} at next-to-leading order (NLO) accuracy in QCD. The Higgs boson decay, and the parton showering and hadronization processes, were implemented by Pythia 8.235~\cite{Sjostrand:2014zea} with the CTEQ6L1 parton distribution function set~\cite{Pumplin:2002vw}. A total of seven million events were generated. For the study, events were required to have at least two photons, each of which should have a transverse energy (\et) greater than 20 GeV and an absolute value of pseudorapidity ($\eta$) of less than 2.5.

\subsection{Detector Response}
For a collider observable $X$, we express its reconstructed value, \xr, as the sum of its true value, \xt, and a term, \dx,  resulting from the experimental effects in the particle detection and reconstruction: $\xr = \xt + \dx$. In this study, we define \dx as a random variable representing the \emph{detector response} of observable $X$. For an ensemble of $X$ measurements, the distribution of its detector response \dx can be modeled by a location-scale family probability density function, $f(\Delta_{X}, \mu(\theta), \sigma(\theta))$, where $\mu$ and $\sigma$ are the location and scale of the function and $\theta$ denotes the dependencies of $\mu$ and $\sigma$.

In lieu of a detector simulation, we can create a proxy of \xr for a given \xt by randomly sampling the detector response function $f(\Delta_{X}, \mu, \sigma)$ and deriving \xr from \xt + \dx. We used this technique to create detector response and reconstruction-level observables that are considered as targets for the \cnf\ model.

\subsection{Experimental Effects in Photon Detection and Reconstruction}
\label{sec:expeffects}
Collider experiments measure photons with an electromagnetic calorimeter (ECAL). For example, the ECAL at the ATLAS experiment is a LAr sampling calorimeter that uses lead/stainless steel as absorbing material and liquid Argon as sampling material~\cite{ATLAS:2008xda}; the CMS experiment has a homogeneous ECAL constructed with Lead-Tungstate crystals~\cite{CMS:2008xjf}. The two experiments, adopting complementary calorimeter technologies,  achieve similar photon detection and reconstruction performances. 
Both used the Crystal Ball or double-sided Crystal Ball functions to model the detector response of photon energy measurements~\cite{ATLAS:2018krz,CMS:2020uim}. Such functions include a Gaussian function to model the core part of the detector response distribution, and power-law functions to model the tails. Various instrumentation effects, such as photon conversions in materials upstream of the calorimeter, the presence of inactive materials in the calorimeter, energy leakage, etc., can introduce a small low energy tail in the energy detector response distribution. For other observables such as the pseudo rapidity $\eta$ and azimuthal angle $\phi$, the detector response may be modeled by a Gaussian function.

At the LHC experiments, multiple proton collisions occur during the same bunch crossing, and this phenomenon is known as \emph{pile up}. The extent of pile up is quantified by the average number of proton interactions per bunch crossing, $\mu$, which has a mean value greater than 30 for the 2017-2018 data-taking periods of ATLAS and CMS experiments~\cite{ATLAS:2019fst,CMS:2020ebo}. Contributions from pile-up collisions deteriorate the measurements of particles arising from the primary collision. As a result, the detector response also depends on $\mu$.

The correlation between measurements of various particles within the same collision event also needs to be considered. For instance, when determining the photon pseudo-rapidity, the collision event primary vertex is used as the photon origin, leading to correlations in the pseudo-rapidity measurements of photons in the same event. The use of pile-up suppression techniques in collider experiments also results in correlations between measurements of different photons, because both measurements receive corrections related to the global energy density of the same collision event.

For an LHC experiment, the reconstruction efficiency of an electromagnetic shower is typically beyond 99\%~\cite{ATLAS:2019qmc}. In data analysis, additional requirements such as photon identification and isolation criteria may be further introduced. The selection efficiency of photons is dependent on its transverse energy and pseudorapidity. As a proof of principle, we do not consider the photon selection efficiency as part of the detector response model. Nonetheless, one could extend the study to incorporate effects of photon identification and isolation criteria. 


\subsection{Parameterization}
\label{sec:para}
In this study, we consider the following photon observables: the transverse energy (\et), the energy projection to the plane perpendicular to the beam axis, the pseudorapidity ($\eta$), and the azimuthal angle ($\phi$). Given these observables for each of the two photons in an \Hgg\ event, we can reconstruct the four momentum of the diphoton system, which is a proxy for the Higgs boson.

Resolutions of these photon observables vary as a function of its truth-level transverse energy and pseudo-rapidity and the event pile-up \m. Specifically, for each photon observable, the photon resolution dependencies are parameterized as follows:
\begin{gather}
    R_{\et}(E_{\text{T,true}}, \eta_{\text{true}}, \mu) = 1.5 \times R_{\et}(E_{\text{T,true}}) \cdot R_{\et}(\eta_{\text{true}}) \cdot R_{\et}(\mu) \\
    R_{\eta}(E_{\text{T,true}}, \eta_{\text{true}}, \mu) = 0.0005 \times R_{\eta}(E_{\text{T,true}}) \cdot R_{\eta}(\eta_{\text{true}}) \cdot R_{\eta}(\mu) \\
    R_{\phi}(E_{\text{T,true}}, \eta_{\text{true}}, \mu) = 0.0003 \times R_{\phi}(E_{\text{T,true}}) \cdot R_{\phi}(\eta_{\text{true}}) \cdot R_{\phi}(\mu)
    \label{eq:resolutions}
\end{gather}
where the resolution's dependencies are modeled separately by fourth-order polynomials $R_{X}(\theta)$ where $X$ represents a photon observable, and $\theta$ are truth-level variables on which photon resolutions depend. The constants in the resolution functions are roughly corresponding to the best resolution values in the parameterization, which are chosen to be compatible with numbers published by the ATLAS experiment~\cite{ATLAS:2018krz}. The polynomial parameterization is given in the Appendix. Figure~\ref{fig:detector_reponse} shows the resolution of measurements of photon kinematic observables \et, $\eta$, and $\phi$, as functions of true values of photon \et and $\eta$, as well as pile-up \m.

\subsection{Scenarios}
\label{sec:scenarios}
We test three different detector response scenarios, which differ in their definitions of the detector response function and the scheme for correlating detector responses between photons. These scenarios are detailed as follows:

\paragraph{Baseline} The photon detector response function $f_{X}(\dx,\mu(\theta),\sigma(\theta))$ is a normal distribution with a mean of zero and width of $R_{X}(\et, \eta, \mu)$. The use of a normal distribution to model the detector response is a simplification in the case of the energy detector response, where a small asymmetry towards lower energy is present due to various instrumentation effects. For each of the two photons in the event, its detector response \dx was sampled from this normal distribution with true values of photon \et, $\eta$, and the event \m as input. The event \m, shared by two photons, was randomly sampled from a uniform distribution between 0 and 40. In the baseline scenario, the detector responses are independent between photons and are normally distributed.

\paragraph{Correlation} We generated detector responses for the two photons in a correlated manner. For a given observable $X$, we used the procedure described in the baseline scenario to create  independent detector responses for the two photons (denoted as $\Delta_{X,1}$ and $\Delta_{X,2}$ respectively). The ordering of the photon is not critical, and we choose to order photons by their transverse energies. To introduce a correlation between the detector responses of two photons, we redefined the detector response for the second photon as follows:

\begin{gather} 
    \Delta_{X,2}^{\mathrm{redefined}} = \rho^2 \Delta_{X,1} + \sqrt{1-\rho^2} \Delta_{X,2} 
\end{gather} 
where the parameter $\rho$ controls the correlation. This parameterization does not aim to replicate any physical correlation scenario between the two photon detector responses. Instead, its purpose is to introduce a controlled correlation in the detector responses, thereby testing the performance of the normalizing flow method. To validate whether our generative model accurately captured the correlation, we created two target samples of events, setting the correlation parameter $\rho$ to either 1.0 or 0.5.

\paragraph{Asymmetric detector response} Due to instrumentation effects, the detector response distribution could be asymmetric, e.g., the ATLAS and CMS experiments use Crystal Ball or double-sided Crystal Ball functions to model their energy detector response. To emulate this behavior in a simplified approach, we define the detector response function as a linear combination of two normal distributions. The core of this detector response function is identical to the normal distribution defined in the baseline scenario.The tail is modeled as a normal distribution with a broader width and a mean that is shifted downwards relative to the central normal distribution's mean. Specifically, the width of the tail distribution is three times that of the central distribution, and its mean is lower by three standard deviations from the mean of the central distribution. These core and tail normal distributions are combined with respective weights of 84\% and 16\%. These asymmetric detector response functions are shown in Figure~\ref{fig:asym_resolution}. The detector responses are drawn independently between two photons.

\section{\label{sec:architecture}Model Architecture}
In this work, a \cnf\ model is trained to learn the transformation that maps a multi-dimensional normal base density distribution onto a target vector $\vec{x}$. This target vector comprises the detector responses of six photon kinematic variables, namely, the \et, $\eta$, and $\phi$ for each of the two photons, making the base density distribution six-dimensional. The conditional vector $\vec{c}$ provided to the \cnf\ model includes the pileup condition $\mu$ and the particle-level kinematic variables \xt, where $X \in \{ E_\mathrm{T}^{\gamma1}, E_\mathrm{T}^{\gamma2}, \eta^{\gamma1}, \eta^{\gamma2} \}$. Superscripts $\gamma1$ and $\gamma2$ denote the two distinct photons. The azimuthal angle $\phi$ is excluded from the conditional input, as collider detectors like ATLAS and CMS exhibit symmetry in $\phi$, resulting in uniform performance in that dimension.

In the inference step, the \cnf model transforms the six-dimensional base density distribution into the six-dimensional distribution of the target $\vec{x}$. The features used as conditional inputs, along with those constituting the target $\vec{x}$, are summarized in Table~\ref{tab:features}.

\begin{table}[]
    \centering
        \begin{tabular}{cc}
        \hline\hline
         Conditional Features & Target Features \\
        \hline

\hline
$E_\mathrm{T}^{\gamma1}$, $E_\mathrm{T}^{\gamma2}$ & $\Delta_{E_\mathrm{T}^{\gamma1}},\Delta_{E_\mathrm{T}^{\gamma2}}$ \\
$\eta^{\gamma1}$, $\eta^{\gamma2}$ & $\Delta_{\eta^{\gamma1}}, \Delta_{\eta^{\gamma2}}$ \\
pile-up $\mu$ & $\Delta_{\phi^{\gamma1}}, \Delta_{\phi^{\gamma2}}$ \\ \hline
        \end{tabular}
    \caption{A summary of the conditional features, and input and output features of the model.}
    \label{tab:features}
\end{table}

The target detector responses of the photon \et, $\eta$, and $\phi$ are scaled to be within $[-1, 1]$. Accordingly, a $\tanh$ bijection is added as the last bijection in the \cnf to ensure the output detector responses are also within $[-1, 1]$. Events that render an absolute value of the scaled detector response above one were discarded in the study. The fraction of rejected events is negligible for the \emph{baseline} scenario and the \emph{correlation} scenario, and it is about 8\% in the \emph{asymmetric detector effect} scenario.
The data sample was split into 80\% for training, 10\% for validation, and 10\% for testing. The validation sample was used to tune the hyperparameters of the model, and the testing sample was used to study the performance of the \cnf model.

The hyperparameters of the \cnf are described as follows. First, the base density distribution, $\pi(\vec{z})$, is chosen to be a multivariate normal distribution, motivated by the overall similarity between detector response distributions and normal distributions. Second, the MLPs inside each MAF module consist of two layers of dense networks with a layer size of 128 and a ReLU activation function~\cite{nair2010rectified}. When we increased the layer size or the number of layers by a factor of two, no significant improvement was observed. Third, we used ten bijection blocks as a result of a trade-off between computational expense and model complexity. Increasing the number of bijection blocks to 20 did not result in any performance improvement compared to the nominal setup of ten bijection blocks. The model might have gained additional improvement if additional training epochs were pursued. Fourth, instead of using a constant learning rate, we employed the Adam~\cite{kingma2015adam} optimizer with a learning rate scheduler that decays the learning rate from $10^{-3}$ to $10^{-5}$ following a power-law distribution; doing so smoothed the training loss distribution and boosted the performance.

\begin{figure}[htb]
    \centering
    \includegraphics[width=0.95\textwidth]{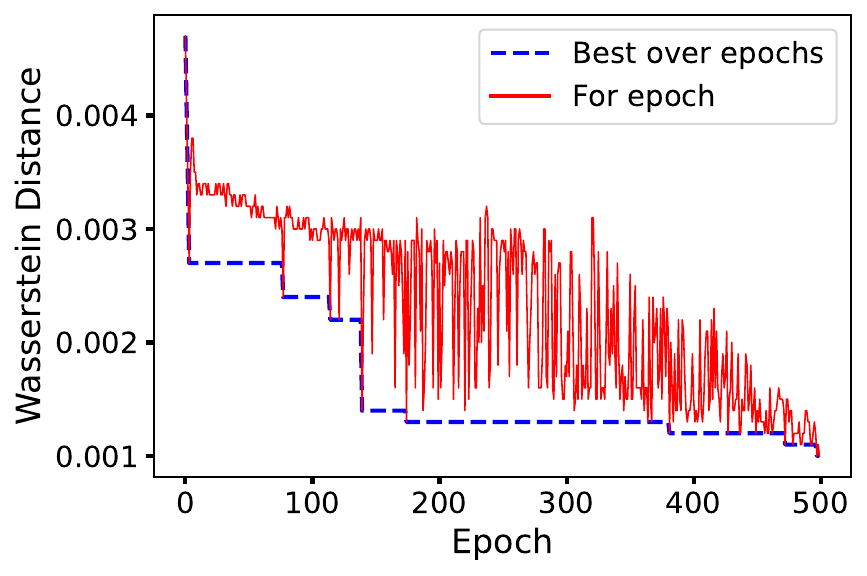}
    \caption{The mean Wasserstein Distance (orange) and the minimum Wasserstein distance (blue) as a function of the training epochs for the baseline scenario. These quantities were evaluated on the validation sample.  }
    \label{fig:loss_vs_epoch}
\end{figure}

All models were trained with 500 epochs and the best model is chosen for testing. We chose to use the ``Wasserstein Distance'' ($WD$)~\cite{villani2008optimal}, a measure of the distance between two probability distributions, to monitor the model performance during the course of training. The $WD$ for a given variable is evaluated between its target and generated distributions. We define the mean Wasserstein Distance, $\overline{WD}$, as the arithmetic mean of the $WD$ values for the six photon detector response variables. After each epoch, the $\overline{WD}$ is evaluated using the validation sample. Figure~\ref{fig:loss_vs_epoch} shows the $\overline{WD}$ for each epoch and the minimum $\overline{WD}$ up to that epoch, as evaluated on the validation sample. The model that yields the minimum $\overline{WD}$ is selected for our study.   

\section{\label{sec:results}Results} 

For each scenario outlined in Sec.~\ref{sec:scenarios},  we trained a separate \cnf model. We then applied the trained model to the test samples and computed reconstruction-level photon kinematic variables using generated detector responses. The sample where the reconstruction-level variables are created from the smearing technique is referred to as the target sample, and the sample where the reconstruction-level variables are calculated from the \cnf generated detector responses is referred to as the \cnf sample.

\paragraph{Baseline scenario}

\begin{figure}[htb]
    \centering
    \includegraphics[width=0.95\textwidth]{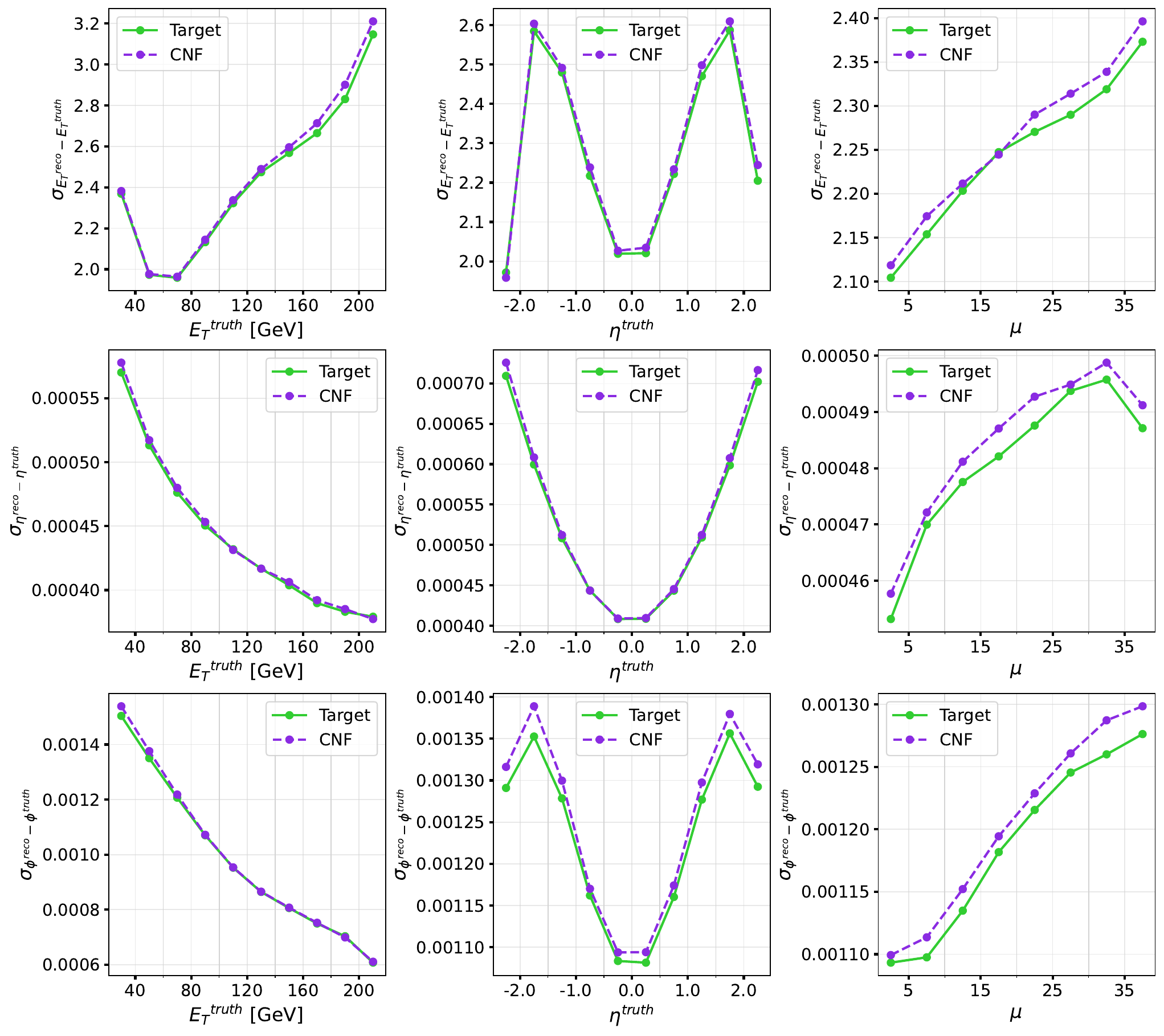}
    \caption{Target and generated photon resolutions $\sigma$ for photon kinematic variables \et, $\eta$, and $\phi$. The resolutions are shown as functions of the true values of photon \et and $\eta$, and the event pile-up \m. The blue (orange) entries represent the target (generated) quantities. The target resolutions corresponding to the \emph{baseline} parameterization presented in Section~\ref{sec:para}.}
    \label{fig:detector_reponse}
\end{figure}

\begin{figure}
    \centering
    \includegraphics[width=0.95\textwidth]{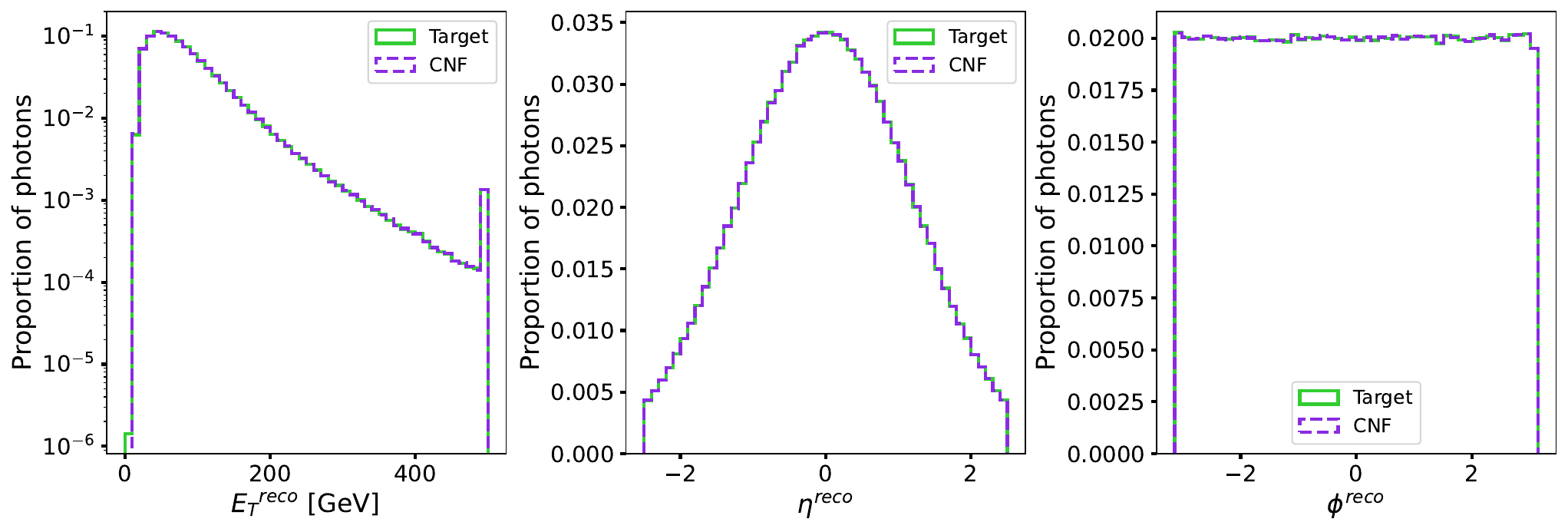}
    \caption{Normalized distributions of photon kinematic variables for the target (blue) and generated (orange) samples. The last bin in the \et distribution contains the overflow entries.}
    \label{fig:photon_reco}
\end{figure}

To quantify the extent in which the \cnf learns the detector response accurately, we calculated the detector resolutions of photons as a function of photon four momenta at the particle level. The detector resolution is defined as the width of the core of the detector response distribution, \dx. Figure~\ref{fig:detector_reponse} shows a good agreement between the target detector resolutions and the \cnf learned ones as functions of photon \et and $\eta$ and event $\mu$. The largest discrepancy is less than 5\%. Figure~\ref{fig:photon_reco} shows the comparison of the target and learned distributions for photon \et, $\eta$, and $\phi$ at the detector level. A good agreement is observed in all distributions. In regions where the statistics of simulated events are low, such as the high \et region, the performance of \cnf would benefit from more simulation events in future studies. 

\begin{figure}[htb]
    \centering
    \includegraphics[width=0.45\textwidth]{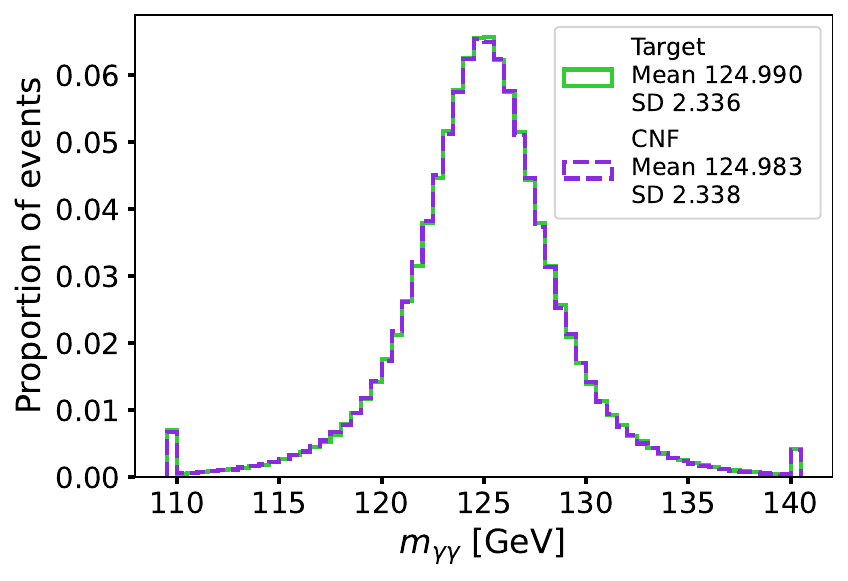}
    \includegraphics[width=0.45\textwidth]{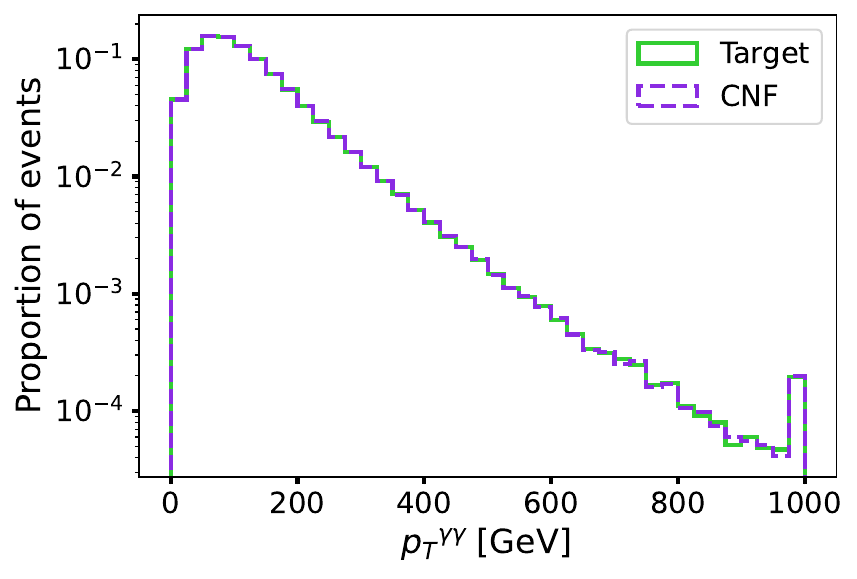}
    \caption{Comparison of the detector-level invariant mass, $m_{\gamma\gamma}$, and transverse momentum of the diphoton system between the target distribution and the \cnf-generated distribution. The two spikes at the edges of the diphoton invariant mass distribution arise from the inclusion of  overflow and underflow entries. The mean and the standard deviation (SD) of the invariant mass distribution are calculated in the mass range of [120, 130] GeV.}
    \label{fig:diphoton}
\end{figure}

We also calculated the invariant mass and transverse momentum of the diphoton system using the target sample and the \cnf\ sample. Figure~\ref{fig:diphoton} shows the comparison of their distributions. The mean and standard deviation values in the diphoton invariant mass distribution from the \cnf\ sample are in agreement with those from the target sample within the statistical precision. For the diphoton transverse momentum distribution, an agreement between the target and the \cnf\ samples is seen across the full range.  

\FloatBarrier

\begin{figure}
    \centering
    \includegraphics[width=0.95\textwidth]{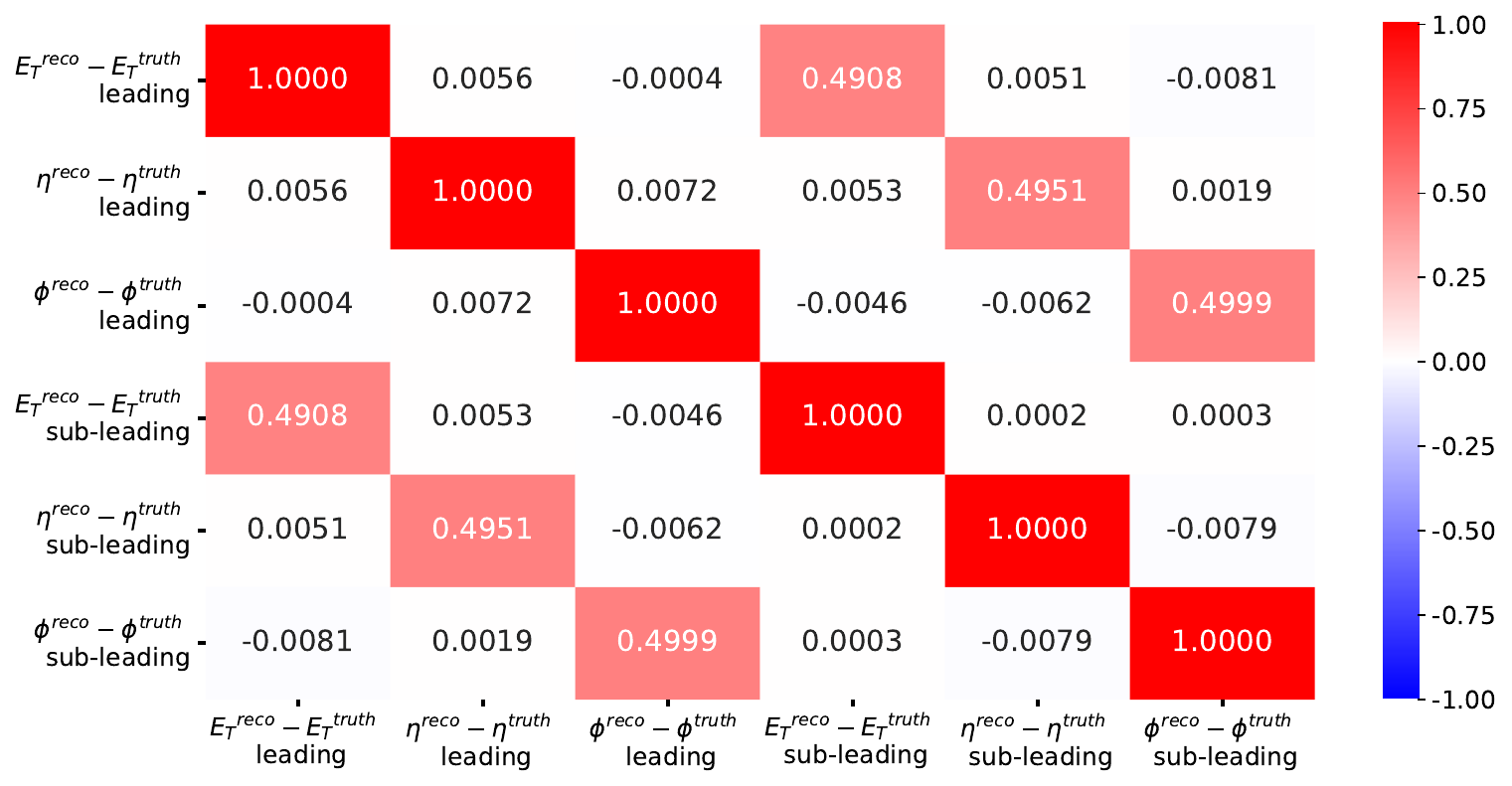}
    \caption{Correlation map of the kinematic resolution of the leading and sub-leading photons for a correlation coefficient of $\rho = 0.5$. Similar performance is observed for the $\rho = 1.0$. In the target correlation map, all off-diagonal entries except those designed to be 50\% correlated are zeros. }
    \label{fig:correlation}
\end{figure}

\paragraph{Correlation scenario}
Two sets of target samples were generated, with the correlation parameter $\rho$ set to 0.5 and 1.0. The \cnf\ model was trained separately for these two samples. Detector responses were generated for the six measurements in the event. Their correlation matrix is shown in Figure~\ref{fig:correlation} using $\rho = 0.5$ sample. The built-in correlation of $\rho = 0.5$ was accurately reproduced. The same performance was also achieved in the case of $\rho = 1.0$. These tests indicate that the \cnf model can accurately reproduce the correlations between the two photons that were built into the measurements. 

\begin{figure}
    \centering
    \includegraphics[width=0.95\textwidth]{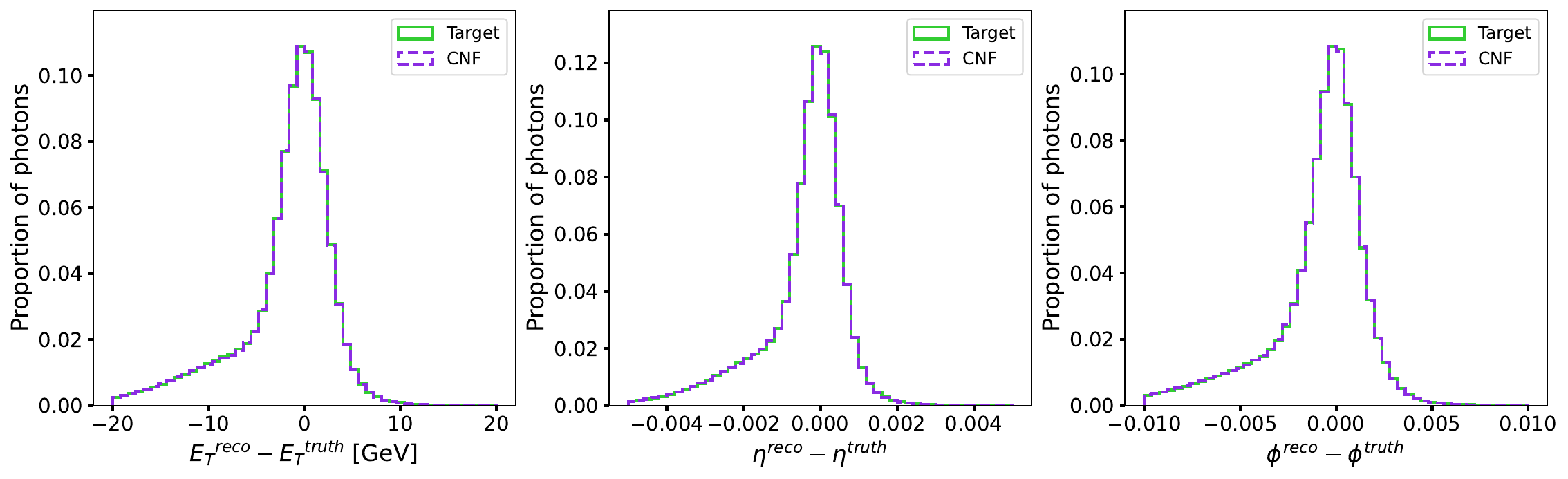}
    \caption{Distributions of detector responses for photon kinematic variables \et, $\eta$, and $\phi$ are shown for the target sample (blue) and the generated sample (orange), in the asymmetric detector effect scenario.}
    \label{fig:asym_resolution}
\end{figure}

\paragraph{Asymmetric detector responses scenario} 
Figure~\ref{fig:asym_resolution} shows the target and generated detector response distributions. The asymmetric tails in various detector response distributions are reproduced by the \cnf model.
\FloatBarrier

\section{Discussion}
\label{sec:discussion}
The studies presented here demonstrate the potential of using normalizing flow-based generative machine learning to model detector responses. In the baseline scenario, we have shown that complex dependencies of detector response on multiple variables can be effectively learned by the \cnf\ model. The correlation and asymmetric detector response scenarios further illustrate the relative advantages of the generative approach compared to smearing-based techniques. Existing smearing-based techniques tend to fall short when dependencies of the detector response on other variables cannot be parameterized analytically or when such parameterization is not straightforward. Moreover, the smearing approach becomes more complex when modeling asymmetric or other irregular detector responses.

For future work, it would be interesting to explore how the performance of the \cnf\ model depends on the implementation of normalizing flows, the choice of base density function, and the model's hyperparameters. For instance, starting with a base density distribution inspired by a known simulation sample, where the detector response already bears some similarities to the target response, might expedite the training process and potentially enhance the \cnf\ model's performance.

The \cnf\ model can be extended to cover other types of detector responses. For instance, in this work, we did not account for the selection efficiency of photons, which is influenced by the application of identification and isolation criteria. These criteria are dependent on the photon's transverse energy, pseudo-rapidity, and event pile-up ($\mu$). The outcome of the photon identification and isolation selection process can be modeled as a binary random variable with values of 0 and 1, where the mean corresponds to the selection efficiency.

Generally, collision events comprise various particle types, each with distinct detector responses, and their multiplicities can also differ between events. Our model is versatile and can be adapted to create detector responses for events containing a larger number and diversity of particles. For instance, one can expand the model's output to produce detector responses for more than two particles simultaneously or apply this model to generate detector responses for a single particle, then sequentially apply it to particle types within the same event.

Given that one of the goals of the generative machine learning method for detector response modeling is to expedite the simulation process, a comparison with other solutions featuring different technical implementations, such as the one in Ref.~\cite{DiBello:2022rss}, would be beneficial. Additionally, examining various types of detector responses and different correlation scenarios between particles in the final state could provide insight into the strengths and weaknesses of different architectures.

\section{\label{sec:conclusion}Conclusions}
In this study, we have explored the use of generative machine learning, specifically a conditional normalizing flow model (\cnf), as a viable alternative to traditional Monte Carlo simulations for modeling detector responses in LHC experiments. Our \cnf\ model, leveraging the Masked Autoregressive Flows, has demonstrated its effectiveness in capturing complex dependencies and correlations in detector responses, as well as managing asymmetric response scenarios.

Our results indicate that the \cnf\ model can accurately simulate detector responses for various particle observables, significantly reducing the computational load compared to conventional simulation techniques. This is particularly noteworthy in scenarios where traditional smearing techniques are inadequate, such as in situations with complex dependencies or asymmetric responses.

\appendix
\section*{Appendix}

The variation of the resolution is parameterized as $R_{X}(x) = \frac{\sum_{i}^{}{ p_i x^{i}}}{\mathcal{C}}$, where $X$ is the measured quantity, $x$ is the variable on which the resolution depends, $p_i$ is the coefficient of the polynomial, and $\mathcal{C}$ is a normalization constant. These parameter values are given in Table~\ref{tab:parameterization}.

\begin{table}[h]
\centering
\caption{Values of $p_0$, $p_1$, $p_2$, $p_3$, $p_4$, $\mathcal{C}$ .}
\label{tab:parameterization}
\begin{tabular}{ccccccc}
\hline
  & $p_0$ & $p_1$ & $p_2$ & $p_3$ & $p_4$ & $\mathcal{C}$  \\ \hline\hline
$\text{R}_{\et}(\et)$ & 1.81 & -0.56 & 0.28 & -0.044 & 0.0024 & 1.46 \\ \hline
$\text{R}_{\et}(\eta)$ & 1.74 & 1.04 & -0.59 & 0.10 & -0.0057 & 1.64 \\ \hline
$\text{R}_{\et}(\mu)$ & 1.74 & 0.058 & 0.0041 & -0.0031 & 0.00031 & 1.74 \\ \hline
$\text{R}_{\eta}(\et)$ & 0.00048 & -6.7e-5 & 1.5e-5 & -1.7e-6 & 7.8e-8 & 0.00032 \\ \hline
$\text{R}_{\eta}(\eta)$ & 0.00066 & -0.00023 & 5.5e-5 & -6.3e-6 & 3.3e-7 & 0.00030 \\ \hline
$\text{R}_{\eta}(\mu)$ & 0.00033 & 2.5e-5 & -9.7e-6 & 2.0e-6 & -1.5e-7 & 0.00033 \\ \hline
$\text{R}_{\phi}(\et)$ & 0.0014 & -0.00021 & 1.0e-5 & 1.6e-6 & -1.6e-7 & 0.00054 \\ \hline
$\text{R}_{\phi}(\eta)$ & 0.00091 & 0.00025 & -0.00016 & 3.0e-5 & -1.6e-6 & 0.00078 \\ \hline
$\text{R}_{\phi}(\mu)$ & 0.00076 & -9.2e-6 & 2.1e-5 & -3.6e-6 & 1.9e-7 & 0.00076 \\ \hline

\end{tabular}
\end{table}
\acknowledgments
This work is supported by the U.S.~Department of Energy (DOE), Office of Science (SC) under contract DE-AC02-05CH11231 and by the DOE award DE-SC0023718. This research used resources of the National Energy Research Scientific Computing Center (NERSC), a DOE SC User Facility operated under Contract No. DE-AC02-05CH11231.

\bibliographystyle{JHEP}
\bibliography{biblio,generative}

\end{document}